# Layered Security Guidance for Data Asset Management in Additive Manufacturing

**Fahad Ali Milaat**
Engineering Laboratory,
Natl. Inst. of Standards and Technology,
Gaithersburg, MD 20899-8260, USA
email: fahad.milaat@nist.gov

**Joshua Lubell**[1]
Engineering Laboratory,
Natl. Inst. of Standards and Technology,
Gaithersburg, MD 20899-8260, USA
email: joshua.lubell@nist.gov

*Manufacturing industries are increasingly adopting additive manufacturing (AM) technologies to produce functional parts in critical systems. However, the inherent complexity of both AM designs and AM processes render them attractive targets for cyber-attacks. Risk-based Information Technology (IT) and Operational Technology (OT) security guidance standards are useful resources for AM security practitioners, but the guidelines they provide are insufficient without additional AM-specific revisions. Therefore, a structured layering approach is needed to efficiently integrate these revisions with preexisting IT and OT security guidance standards. To implement such an approach, this paper proposes leveraging the National Institute of Standards and Technology's Cybersecurity Framework (CSF) to develop layered, risk-based guidance for fulfilling specific security outcomes. It begins with an in-depth literature review that reveals the importance of AM data and asset management to risk-based security. Next, this paper adopts the CSF asset identification and management security outcomes as an example for providing AM-specific guidance and identifies the AM geometry and process definitions to aid manufacturers in mapping data flows and documenting processes. Finally, this paper uses the Open Security Controls Assessment Language to integrate the AM-specific guidance together with existing IT and OT security guidance in a rigorous and traceable manner. This paper's contribution is to show how a risk-based layered approach enables the authoring, publishing, and management of AM-specific security guidance that is currently lacking. The authors believe implementation of the layered approach would result in value-added, non-redundant security guidance for AM that is consistent with the preexisting guidance.*

## 1 Introduction

Additive manufacturing, also known as 3D printing, is the collection of technologies and processes for joining materials to build products layer upon layer from 3D model data [1]. Additive manufacturing provides technological and economical means for developing agile, sustainable, and cost-effective business models [2]. In recent years, additive manufacturing adoption has increased across multiple manufacturing sectors. For instance, the 2022 Wohlers Report [3] noted that the Boeing Company is flying over 70,000 additive manufactured parts on aircraft and satellites, and the U.S. Food and Drug Administration (FDA) cleared over 250 medical devices produced by additive processes. However, the deployment of additive manufactured parts in safety-critical systems make them prime targets for cybersecurity attacks [4].

Additive manufacturing security can be considered a special case of cyber-physical system security, where an attack may not only disrupt the printing machine equipment, but also introduce defects to the manufactured 3D part [5]. Furthermore, the security of additive manufacturing extends beyond cybersecurity challenges that could be addressed through information technology (IT) security guidelines such as those for cryptographic standards [6], and operational technology (OT) security guidance such as those for industrial internet of things [7]. The reason for such distinctive challenges is related to the very nature of additive technologies and their dependencies on digital information in guiding the physical construction, and qualification of a 3D part. However, the sole reliance on fundamental concepts in information security, such as confidentiality, integrity, and availability, could be detrimental to the development of methods and protocols that are specifically tailored for additive manufacturing security [8]. In [9], Yampolskiy et al. developed two categories of additive manufacturing-based taxonomies of attack targets. The first focused on intellectual property (IP) theft of 3D computer-aided designs (CAD), process parameters, and post-processing specifications. The second focused on sabotage attacks that inflict damages to manufactured parts, machine equipment, and environment.

Successful exploitation of attack targets could have disruptive consequences for a multitude of additive manufacturing processes. Such targets include 1) introducing deceptively subtle defects in a part through process parameter manipulation [10], and 2) compromising a network of stakeholders in the manufacturing supply chain [11]. When exploring the open literature on additive manufacturing security, only a handful of solutions considered the complexity of their platform ecosystem [12], or developed a general-purpose framework for cyber-attack detection [13]. Nevertheless, a recent survey of the America Makes additive manufacturing community showed that only 41% of the participating organizations dedicated security programs for their additive value chains [14]. Therefore, there is a need for additive manufacturing-specific standards that are consistent with preexisting guidance.

This paper applies a layered guidance approach for additive manufacturing security standardization. Specifically, the layering method combines guidance from existing risk management frameworks and other sources with new guidance specific to additive manufacturing such that guidance is traceable to its source and is non-redundant. This paper's layered approach uses the National Institute of Standards and Technology's Cybersecurity Framework (CSF) [15] as its foundation. The CSF defines "risk" as "a measure of the extent to which an entity is threatened by a potential circumstance or event, and typically a function of: (i) the adverse impacts that would arise if the circumstance or event occurs; and (ii) the likelihood of occurrence." The CSF structures security requirements as a hierarchy of outcomes useful for communicating an organization's current or desired security posture. At the top level are five broad risk management functions: IDENTIFY (ID), PROTECT (PR), DETECT (DE), RESPOND (RS), and RECOVER (RC). A set of security outcome categories subdivide each function. A set of outcome subcategories further divides each category. Outcomes may include informative references that illustrate methods to achieve them. Informative references are also used to align CSF outcomes with related concepts defined in other standardized security frameworks, such as ISO/IEC 27001 [16].

---

[1]Corresponding Author.
September 28, 2023

To help support achievement of CSF outcomes, organizations may select a subset of the CSF categories and subcategories and supplement them with additional guidance that is targeted to the operational capabilities of the system or process. Such an adaptation is called a CSF *profile*. Section 3 provides examples of both additional guidance specific to additive manufacturing and a reference to resources helpful for achieving outcomes within the CSF Asset Management (ID.AM) category, chosen because asset management is critical to risk-based security. Multiple research works, for example, Kure and Islam [17] and Lyu et al. [18], assert that identification and analysis of assets should be the initial step of any approach to risk management. Furthermore, ID.AM illustrates the benefits of the layered guidance approach since, as Sec. 3 and Sec. 4 will show, there are instances where additive manufacturing-specific guidance is needed to supplement and (in one case) replace existing OT guidance.

To demonstrate how the proposed layered approach could be automated, Sec. 4 shows how the Open Security Controls Assessment Language (OSCAL) [19] can be used to integrate additive manufacturing-specific data asset management guidance with preexisting IT and OT security guidance. The integration will yield the desired composition of CSF outcomes with additive manufacturing-specific requirements. This paper's contribution is to show how a risk-based layered approach enables the authoring, publishing, and management of additive manufacturing-specific security guidance that is currently lacking. This layered guidance leverages preexisting IT and OT security standards.

The paper is structured as follows. Section 2 reviews the body of published additive manufacturing security research. Section 3 first describes a motivating example within the scope of the CSF Asset Management (ID.AM) outcome category that illustrates the layered approach. Then it discusses data representations for additive manufacturing being standardized within ISO that could support some ID.AM outcomes. Section 4 shows how the data models of OSCAL support automation of the proposed layered approach. Finally, Section 5 concludes this paper.

## 2  Literature Review

This section reviews recent research literature that focuses on the security of additive manufacturing. Following the additive manufacturing attack taxonomy proposed by Yampolskiy et al. [9], the first part will cover intellectual property theft attacks and their countermeasures, and the second part will cover sabotage attacks and their countermeasures. The third part will discuss emerging attacks and the challenges associated with protecting against them.

**2.1 Intellectual Property Theft.** IP protection is an essential element for the success of additive manufacturing business models [20]. In the additive manufacturing domain, IP could be represented in geometric part definitions, process parameters and plans, or machine controls and measurements. However, as more advanced and less expensive additive machines enter global markets, the feasibility of IP theft with additive technologies become apparent [21]. Nevertheless, the emergence of side-channel attacks in additive manufacturing, such as the acoustic attack first proposed by Al-Faruque et. al [22], questions the sufficiency of conventional IT security protocols along the additive process chain. For instance, Gatlin et al. [23] investigated the feasibility of additive manufactured parts reconstruction through a novel side-channel attack, which draws current directly from positional actuators in an additive manufacturing machine. The proposed attack envisioned a Man-at-the-End threat scenario, where an additive manufacturing machine could be instrumented by the attacker at will to deduce the current amplitudes of positional actuators. The attack does so without compromising cyber security measures or gaining access to the build command files. Despite the reconstruction limitations described by the authors, the proposed side-channel attack was able to achieve 99 % reconstruction accuracy of additive manufactured parts. This result highlights the need for alternatives to encryption in some scenarios.

To counter side-channel attacks in additive manufacturing, Brandman et al. [24] proposed a physical hash that couples digital information with the manufactured part through an air-gapped side-channel measurement system. Printed alongside the actual additive part, the quick response (QR)-formed physical hash stores a hash string that represents the side-channel measurements of the designed part. During manufacturing, side-channel measurements are collected in-situ from sensors within an additive system, hashed and compared to the printed QR code, which is then scanned to determine conformance to specifications of the designed part. This method enables quality specifications to be transmitted to the air-gapped side-channel measurement system in a secure manner, without exposing IP information or process controls. Liang et al. [25] proposed an optical side-channel attack on an additive machine, which uses camera footage overlooking the build as input into a deep neural network (DNN) model. This DNN model can be used to predict the coordinates of a 3D printing head and recover IP of the fabricated part. To mitigate this side-channel attack, the authors proposed an optical noise injection method, where a crafted video file is optically projected onto the build area of the 3D printer to confuse the adversarial recovery of the print design. Although the proposed mitigation method could easily defeat a naive attacker who is unaware of the injected optical noise in the camera footage, the authors developed robust noise generation algorithms against an advanced attacker.

**2.2 Sabotage Attacks.** A sabotage attack is defined as any intentional action taken to damage, modify, or compromise an asset, whether physical or virtual, with the expectation of harmful outcomes. In the additive manufacturing domain, sabotage attacks could target the manufactured part, the machine equipment, or the machine environment [9]. The following review covers each form of additive manufacturing sabotage attack.

*2.2.1 Targeting additive manufacturing parts.* Belikovetsky et al. [26] presented a novel sabotage attack that reduced the fatigue life of an additively manufactured propeller of an aerial quad-copter. This sabotage attack infiltrated the additive manufacturing workflow by compromising an Internet-connected PC that controlled an additive machine. It did so by retrieving and manipulating the propeller's files, and replacing the original files with the modified ones. By targeting the propeller's design and process files, the authors demonstrated a complete sabotage attack chain that resulted in breaking the propeller mid-flight and destroying the quad-copter during operation. In [27], Carrion et al. proposed a sabotage attack that selectively modified laser-based powder bed fusion (L-PBF) process parameters that were used to manufacture two stainless-steel-alloy parts. For this attack, laser power parameters were altered in specified cross-sections of a part design to imitate localized layer thickness modifications due to unmolten powder. Tensile strength degradations were verified using non-destructive techniques (NDT) at the targeted cross-sections of the compromised parts. Parker et al. [28] investigated security threats that sabotaged additively manufactured molds, causing indirect impacts on parts being produced using die casting techniques. In a systematic analysis, the authors categorized sabotage attacks and their impacts on mold design and manufacturing, and compared their characteristics to attacks that targeted the direct additive manufacturing of functional parts.

To evaluate the subtlety of G-code modification in additive manufacturing, Beckwith et al. [29] conducted a case study where a red team introduced subtle defects designed to impact the additive part, and a blue team detected those alterations without access to the original part design. Shi et al. [30] proposed an online side-channel monitoring approach based on long short-term memory (LSTM) autoencoder to detect additive manufacturing alterations caused by cyber-physical attacks. The machine learning (ML)-based approach featured supervised and unsupervised monitoring schemes that utilized side-channel data from an additive machine. Those schemes used that data for feature extraction and detection of fused filament fabrication (FFF) process alterations. Al-Mamun et al. [31] proposed an additive manufacturing process authentication method that applied feature extraction techniques on streamline videos collected from a camera attached to a FFF printing head. The proposed method observed the texture of each layer and constructed a layer-wise texture descriptor tensor, where a Hotelling control chart technique was adopted for the detection of additive process alterations.

*2.2.2 Targeting additive manufacturing equipment.* Considering the threat of additive manufacturing equipment sabotage, Graves et al. [32] investigated cyber-physical attacks aimed at sabotaging the powder delivery system of a metal L-PBF machine. The investigation identified potential manipulations that could produce defects in the additive part, and evaluated the impacts of selected powder delivery system-based attacks experimentally using NDT and destructive testing approaches. Pearce et al. [33] presented a Trojan-based attack named "FLAW3D", which targeted the bootloader component of an additive machine's firmware. The attack replaced the bootloader's interrupt service routines with malicious routines and forced their execution in the main application. This enabled the interception of incoming commands, such as G-code sentences, and the potential for introducing part defects by altering the build process.

In mitigating the threat of additive manufacturing equipment sabotage, Gatlin et al. [34] proposed a sabotage detection approach based on trusted monitoring of the supplied current to individual actuators of an additive machine and the detection of anomalies indicative of sabotage attacks. The proposed approach was evaluated on a Fused Deposition Modeling (FDM) additive machine, where currents of the X, Y, Z axis motors and the filament extrusion motor were monitored for anomalies caused by insertion, deletion and reordering of G-code movement commands.

Other mitigation methods employed ML approaches. In [35], Yu et al. proposed a multi-modal sabotage attack detection system for additive machines. The proposed detection system focused on firmware manipulation attacks, assuming that G-code commands were intact. At first, the system continuously monitored and analyzed multiple side-channels of an FDM machine, such as nozzle speed and temperature, during the additive fabrication of a part. Side-channel observations were then compared to the additive machine's uncompromised control signals for the selected part. For detection, ML models were used to identify unusual analog emissions resulting from potential sabotage attacks. Similarly, Rott and Monroy [36] designed a DNN method that detected sabotage attacks by predicting the power consumption of a FDM machine given the additive part design and the previous power consumption of the uncompromised build. The DNN generated predictions of the average peak AC current supplied to the FDM machine based on interpreted tool-path instructions from G-code commands. Experimental evaluations of the proposed method showed 96 % accuracy in detecting defect injection attacks targeting an additive part.

*2.2.3 Targeting additive manufacturing environment.* The additive manufacturing environment includes not only the physical build chamber, but also but also where post-processing occurs. Compromising the additive manufacturing environment requires in-depth understanding of both the manufacturing technology and the environmental properties that could cause part degradation. Zinner et al. [37] investigated the ability of using the shielding gas flow system of a L-PBF machine to conduct a sabotage attack. The investigation analyzed the shielding gas throughput and its impact on the quality of the manufactured parts. Investigation findings concluded that environmental sabotage attacks could be characterized as probabilistic. This is important because it can not only increase the complexity of accurately mounting sabotage attacks, but also introduce uncertainties that might not be directly attributed to intentional sabotage.

To detect an environment-based sabotage attack, Kurkowski et al. [38] developed an anomaly detection method for L-PBF systems. The method monitored multiple side-channels in the form of print bed movement, laser emittance time, and print chamber temperature without access to toolpath control commands. The collected observations were used to create a baseline of the ground truth additive part fabrication, and detection thresholds were configured based on the L-PBF machine's tolerances and chosen precision. The anomaly detection algorithm compared side-channel observations of the fabricated part against the truth baseline using the specified thresholds. Experimental results yielded 96.9 % accuracy in detecting anomalies caused by active attacks.

**2.3 Emerging Attacks and Mitigation Challenges.** The presented literature review highlights the diversity of vulnerabilities and security threats in each attack target category of additive manufacturing. However, emerging attack categories such as illegal part manufacturing and steganographic attacks further exacerbate the challenge of managing risks in additive applications. In the case of illegal part manufacturing, successful execution could cause an original equipment manufacturer (OEM) severe financial loss and reputation damage. To this end, Yanamandra et al. [39] explored the feasibility of reverse engineering an additive part through reconstructing the original tool-path commands using imaging and ML methods. In [40], Sola et al. reviewed tagging strategies that could be used to identify and fingerprint additive parts. Wei et al. [41] proposed an anti-counterfeiting approach that embedded metallic security features into an L-PBF manufactured part using an ultrasonic selective powder delivery system.

On the other hand, steganographic approaches can hide information inside the digital file of an additive part, or place it directly onto an additive part in the form of 3D QR codes. In [42], Usama and Yaman conducted a review of additive manufacturing information embedding techniques such as 3D QR codes, steganography, and watermarking. Each of these techniques could be implanted into or onto an additive part for tracking, interaction, and authentication purposes. However, the proliferation of data exchanges along the additive manufacturing digital thread create opportunities for an attacker to subvert regular communication channels to hide valuable information such as trade secrets. Yampolskiy et al. [43] explored steganographic attacks where arbitrary information could be covertly hidden into an STL file of an additive part without distorting the geometry of the printed part.

As more and more vulnerabilities and novel attacks are discovered in additive manufacturing, the need for additive-specific security guidance becomes increasingly apparent. This section's literature review has shown that additive manufacturing systems and processes have unique characteristics that not only add new opportunities for attackers, but also may add new ways to detect and respond to such attacks. Therefore, risk-based additive manufacturing security guidance is needed. This guidance must distinguish vulnerabilities and threats specific to the additive ecosystem from conventional cyber-physical systems, which could not only wreak havoc in the cyber (e.g., CAD data) and cyber-physical (e.g., machine instruments) domains, but also result in defective or ille-gally manufactured parts. Furthermore, a robust implementation of a layering guidance approach is essential to ensure that guidance standardization remains synchronized with the ongoing discovery of new attacks and mitigation techniques. The applied layering approach described in the next two sections provides a mechanism for tailoring and extending existing IT and OT security guidance for additive manufacturing.

## 3 Layered Security Guidance for Additive manufacturing

This section describes the layered security guidance approach for additive manufacturing. Furthermore, this section shows how a Cybersecurity Framework (discussed in Sec. 1) additive manufacturing asset management profile can leverage existing guidance for IT and OT asset management. The approach is motivated by the security awareness cycle [4], in which new technology is initially developed without much focus on security. Once industry adoption becomes widespread and security become a greater concern, the new technology tends to inherit security guidance from similar or related technologies. However, as security experts (best case) or bad actors (worst case) discover attacks for which existing security guidance is lacking, the need to update the existing security guidance becomes evident. The literature review in Sec. 2 shows that this is now the case for additive manufacturing.

Fig. 1 shows the applied layered approach. The four layers on the left, each with its own top arrow pointing rightward, all contribute to the additive manufacturing security guidance document on the right. The first two layers are sources of preexisting guidance. One is the full set of CSF category and subcategory outcomes. The other is the OT CSF profile described in the NIST Guide to OT Security (Special Publication 800-82 Revision 3, Sec. 6) [7]. The Guide's CSF OT profile enhances CSF outcomes with two types of guidance. General guidance, although not specific to OT, adds additional context to aid in better understanding of the CSF outcome. OT-specific guidance is intended to address considerations more unique to OT.

The next layer is a profile for additive manufacturing that is meant to be general enough to apply to most additive process technologies. This additive manufacturing profile may inherit all of the outcomes in the OT profile, or it may be scoped to inherit just a subset of the OT profile's CSF outcomes. Likewise, the additive manufacturing profile might inherit some but not all of the OT profile's guidance. Indeed, Sec. 3.1 shows a case where an outcome's OT-specific guidance might not be broadly applicable to additive manufacturing. In addition to additive-specific guidance, an additive manufacturing profile may contain informative references to resources external to the profile. Such resources might provide further guidance in achieving the CSF outcome. Figure 1's additive manufacturing profile layer lists two examples of informative references. The first focuses on attack taxonomies such as Yampolskiy's [9], which can be helpful as part of Risk Assessment, one of six categories within the CSF IDENTIFY function. Risk Assessment (CSF category ID.RA), which depends on the ID.AM outcomes, is also highly important to risk-based security. Although not part of the example presented in Sec. 3.1 and Sec. 4, this paper's concluding section (Sec. 5) suggests risk assessment for additive manufacturing as a topic for future work. The second informative reference is a data model for general additive manufacturing (see Sec. 3.2) that could help in mapping data flows and prioritizing data assets based on their classification, criticality, and business value. Both of these tasks support ID.AM, which is the main focus of this Section's example.

Figure 1's rightmost layer shows a CSF profile specific to L-PBF process technology, which uses high power beams to fuse powder material into layers of scanned patterns to produce parts with complex geometries. Although not discussed further in this paper, such a profile could be developed as an additional layer specializing the general additive manufacturing profile. The specialization could consist of additional process-specific guidance (and removal of any

**Table 1  Asset management (ID.AM) subcategories**

| Identifier | Outcome |
|---|---|
| ID.AM-1 | Physical devices and systems within the organization are inventoried |
| ID.AM-2 | Software platforms and applications within the organization are inventoried |
| ID.AM-3 | Organizational communication and data flows are mapped |
| ID.AM-4 | External information systems are catalogued |
| ID.AM-5 | Resources (e.g., hardware, devices, data, time, personnel, and software) are prioritized based on their classification, criticality, and business value |
| ID.AM-6 | Cybersecurity roles and responsibilities for the entire workforce and third-party stakeholders (e.g., suppliers, customers, partners) are established |

general guidance not applicable to L-PBF). Rather than the general additive manufacturing data model referenced in an additive manufacturing profile, a L-PBF-specific profile might reference a L-PBF data model proposed by Milaat et al. [44] based on the general additive manufacturing data model. In addition to the attack taxonomies referenced in the additive manufacturing profile, a L-PBF profile could reference Malekipour's and El-Mounayri's taxonomic representation of L-PBF-manufactured part defects and their contributing process parameters [10].

**3.1 Additive-specific Asset Management Guidance and Illustration of Layering.** ID.AM's outcome statement is as follows [15]:

> The data, personnel, devices, systems, and facilities that enable the organization to achieve business purposes are identified and managed consistent with their relative importance to organizational objectives and the organization's risk strategy.

Table 1 shows ID.AM's subcategories [15]. The remainder of this section provides a detailed example focusing most heavily on ID.AM-3. This paper stresses ID.AM-3 because ID.AM-3 specifically pertains to data, and additive manufacturing processes are data-intensive. Prior to discussion of ID.AM-3, it is helpful to briefly comment on the additive manufacturing-specific guidance needs of the other five ID.AM subcategories. For ID.AM-1, one could supplement existing guidance with a reminder that inventory management should include the feedstock. For ID.AM-2, additive-specific guidance can state that the software includes all applications and libraries comprising the tool chain for transforming a digital object specification into a 3D-printed part. Such a task might be simplified by incorporating the concept of the Software Bill of Materials (SBOM) [45] to capture these details. ID.AM-4's guidance includes third-party systems and services that an organization relies on, but are managed by external providers. In an outsourced additive manufacturing scenario, these might include a supplier's 3D printer and e-commerce platform. ID.AM-5 might include guidance suggesting that CAD and computer-aided manufacturing (CAM) data could be classified differently. For example, process parameters might be deemed to have higher criticality and business value than the design data.

ID.AM-6 might need additive-specific guidance accounting for the flexible business models additive manufacturing technology enables. For example, one could easily imagine a scenario where a customer in need of 3D printing services would require a provider to take responsibility for protecting the customer's confidential data. On the other hand, Gupta et al. [11] provide an intriguing example of a scenario where a customer might choose *not to* require the provider to ensure confidentiality. In this scenario, the customer is an antique automobile owner in need of a replacement part no longer sold but for which a public domain 3D part model is available. The customer requires that the service provider must 1) assume responsibility for maintaining the integrity of the customer's data and 2) ensure that the printed part meets the design specifications. However, the service provider has no requirement to ensure data confidentiality.

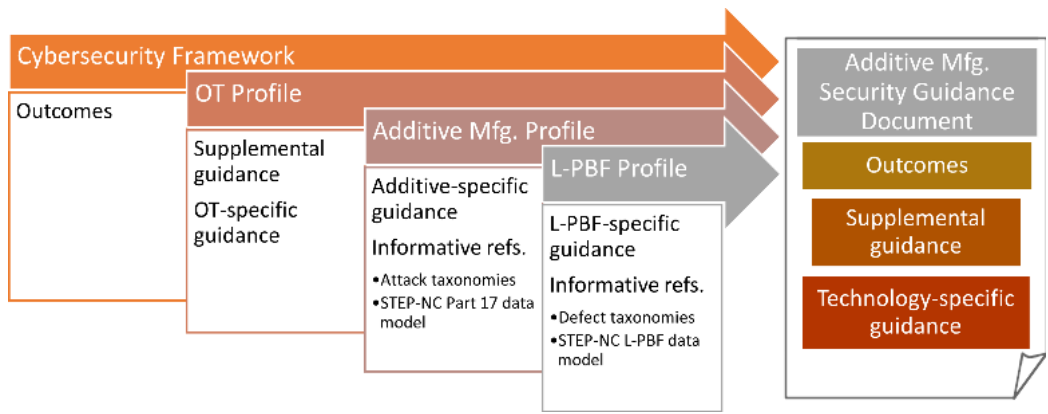

**Fig. 1  Layering additive manufacturing security guidance**

Having briefly discussed ID.AM-1, ID.AM-2, ID.AM-4, ID.AM-5, and (in greater detail) ID.AM-6, this paper now focuses on ID.AM-3. For additive manufacturing, ID.AM-3 is important because it includes not only the flow of data over a network, but also the relationships between the data that model the processes, structures, and properties associated with a part's fabrication. Application of the layered approach to modify ID.AM-3 for additive manufacturing cybersecurity begins by examining the general and OT-specific guidance for ID.AM-3 provided in NIST's Guide to OT Security [7]. The general guidance from the Guide, which supplements ID.AM-3's outcome statement, is as follows (condensed for brevity):

> Data flow diagrams enable a manufacturer to understand the flow of data between networked components. Documenting data flows enables organizations... This information can be leveraged during forensic activities or used for analysis to identify anomalies.

This guidance states that data flow mappings between networked system components are important because the mappings help organizations understand their networks and detect unexpected behavior. Since this guidance is relevant to additive manufacturing, Sec. 4's automation example of the layered approach includes this guidance in a CSF additive manufacturing profile.

The Guide to OT Security's OT-specific guidance for ID.AM-3 is as follows (condensed for brevity):

> Organizations should consider the impact on OT systems from the use of automated data flow mapping tools that use active scanning or require network monitoring tools (e.g., in-line network probes)... Consider using data flow mapping tools that utilize these methods during planned downtime.

This guidance, which applies to data flows across a network, cautions that automated mapping tools could negatively impact information availability and thus hurt system performance. For demonstrating the principles of this paper, additive manufacturing environments are assumed to be sensitive to network scanning, thus requiring tailored approaches for time sensitive processes. However, the guidance may matter less for additive manufacturing systems than other OT systems. Additive manufacturing processes allow for highly complex CAx (computer-aided design, process planning, and manufacturing) data. Compromised network transmission is only one of multiple ways additive manufacturing data can be compromised. Compromise can also result from an attack on any of the software applications that transform the data during the CAx process. These software applications may communicate with one another over a local network, or some could reside on a single computer whose purpose is to control the 3D printer. Attacks on these software applications could exploit software vulnerabilities, or they could be triggered by providing faulty inputs. Thus, for additive manufacturing, a local network-based automated mapping tool is limited in its ability to map all organizational communication and data flows. Therefore, Sec. 4's automation example assumes that ID.AM-3's OT-specific guidance should not be included in a CSF additive manufacturing profile.

Existing information models that represent additive manufacturing processes can be helpful in providing the mapping ID.AM-3 requires. One such information model is provided in ISO 14649-17 [46], also known as part-17 for short. Sec. 3.2 discusses how this information model, augmented with additional proposed data representations of AM process information, can help with implementing ID.AM-3. Coupled with taxonomic representations linking process parameters to part defects [10] or to entities in an additive manufacturing supply chain [11], information models such as part-17 are useful for prioritizing which process data elements are most critical (ID.AM-5) and therefore need the most protection from threats.

A CSF additive manufacturing profile might augment the Guide to OT Security's guidance for ID.AM-3 with guidance something like the following (condensed for brevity):

> Data flow diagrams for AM processes represent the flow of data between networked components, as well as relationships between data elements and their impacts on printed parts. AM provides more freedom in design complexity than other manufacturing technologies. Therefore, AM processes are heavily data-driven. Examples of AM data elements that can be manipulated to adversely affect the manufacturing process include CAD geometry... Examples of adverse impacts include porosity...

**3.2 Additive Manufacturing Data Representations.** The digitization of additive manufacturing information spans the entire manufacturing process chain from part design to fabrication, inspection and qualification. For achieving ID.AM outcomes, organizations should examine the data formats in use to support their digital manufacturing. However, most of the additive manufacturing industry today still use file formats such as STL for geometry tessellation, and G-code when developing process plans and controls. This brings more challenges to the security of additive manufacturing because the development of such file formats was limited in scope (e.g., tessellated geometry but without tolerance) and focused more on functionality rather than security. Furthermore, the flow of information through additive stages necessitates, in most cases, the conversion of data from one format to another. But doing so lacks any additive-specific guidance for mapping data flows to meet asset management requirements. Yet, these conversions increase opportunities for sabotage attacks that seek to introduce part defects by modifying geometry definitions and process parameters.

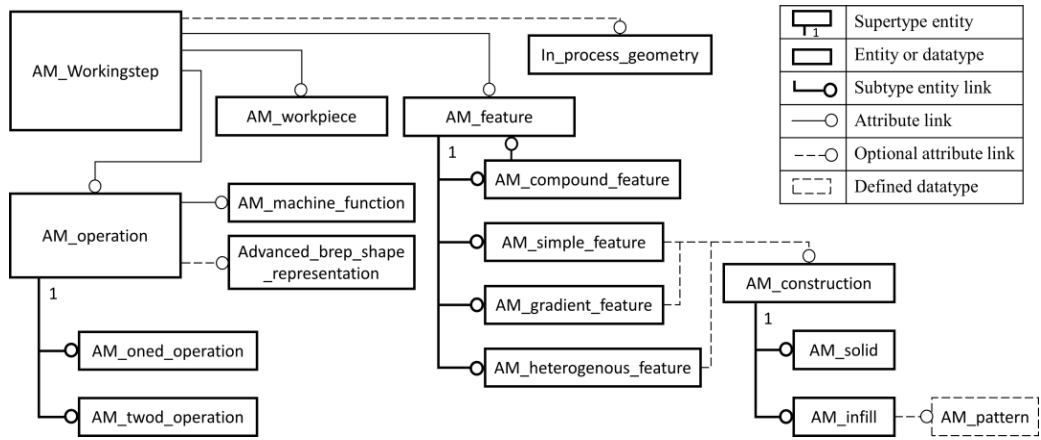

**Fig. 2  ISO 14649-17 standard entities [44]**

Several file formats have been introduced to improve the representation and interoperability of additive manufacturing information such as the Additive Manufacturing File (AMF) [47] and the 3D Manufacturing File (3MF) [48]. Both file formats use Extensible Markup Langauge (XML) schemas to represent tessellated geometries, materials, colors, and orientation of an additive part, with 3MF adding digital signature capabilities for document verification [48]. However, the AMF and 3MF formats do not extend their definitions to include process planning or process controls for additive technologies, and therefore require conversions into another format that enables process plans to be specified. Recent studies have explored the viability of representing the geometry of an additive part and its fabrication process plans in the Standard for the Exchange of Product Model Data with Numerical Control (STEP-NC) [49] [50] [51].

The STEP-NC format is an ISO standard [52] that provides descriptive models for geometry and tolerance, as well as process control and machining models. The standard also facilitates their exchange between software applications in a system-neutral format [53]. Recently, the STEP-NC models were updated in the ISO 14649-17 standard [46], or part-17 for short, with data representations of geometry and processes for general additive manufacturing. Figure 2 shows a schema for data entities of part-17 that are represented in the graphical EXPRESS language [54]. Although not a data flow diagram, the graphical schema is a useful visual aid for efficient evaluation and interpretation of data entities. Here, the "AM_workpiece" describes the 3D model of an additive part, whereas the "AM_workingstep" contains the sequence of operations, or "AM_operation", that are executed on a particular feature of the geometry, or "AM_feature". The detailed definitions and relational structures of STEP-NC models present a viable solution to the ID.AM-3 requirement by establishing additive manufacturing data flows that map the part geometry to the corresponding operation processes in a standardized file format. In this manner, STEP-NC models can provide human readable, graphical illustrations that would assist cybersecurity personnel in intuitively assessing, and effectively responding to, potential incidents. However, the current part-17 standard does not convey technology-specific definitions that are requisite for additive manufacturing process planning and control.

Milaat et al. [44] proposed new STEP-NC compliant data representations for the L-PBF technology. Specifically, new process parameters were introduced to the "AM_operation" entity. These parameters provide process control specifications such as hatch space and inter-layer rotation, technology specifications such as beam power output and scan speed, and scan strategy specifications for stripe and chess patterns, respectively. By expanding STEP-NC definitions to include technology-related process parameters, one could realize the ID.AM-5 requirement through prioritizing the parameters that are critical to the material properties, mechanical properties, and near-net-shape of an additive part.

**3.3 Summary.** Section 3.1 described the layered approach from the perspective of a guidance author with expertise in both additive manufacturing technology and risk-based frameworks such as CSF. A concrete example using ID.AM-3 illustrated the methodology. Section 3.2 covered standardized data representations of additive manufacturing geometry and processing definitions. These standardized representations support ID.AM-3 and ID.AM-5 in an additive manufacturing CSF profile. Section 4 discusses the layered approach from an implementation perspective, i.e., from the viewpoint of developer of productivity tools for authors and maintainers of CSF additive manufacturing profile content. Specifically, Sec. 4 proposes OSCAL as meeting the requirements for representing CSF profiles in a machine-actionable manner.

## 4  Automating the Layered Approach with OSCAL

As the expanded example in Sec. 3 for ID.AM-3 showed, the layered approach requires more than just adding additive manufacturing-specific guidance to preexisting, more general guidance. Sometimes the preexisting guidance may not be suitable for including in an additive manufacturing security guidance standard. Therefore, the layering approach must accommodate both composition (adding new guidance to preexisting guidance) and replacement (deleting preexisting guidance and replacing it with new guidance). Additive manufacturing guidance standards need updating to reflect new advances in process technology. Revisions to guidance inherited from lower layers might also trigger the need for an update. Therefore, the layering approach must also support change management. This implies the need for software tools with interfaces that help to automate the authoring, publishing, and managing of these new guidance documents.

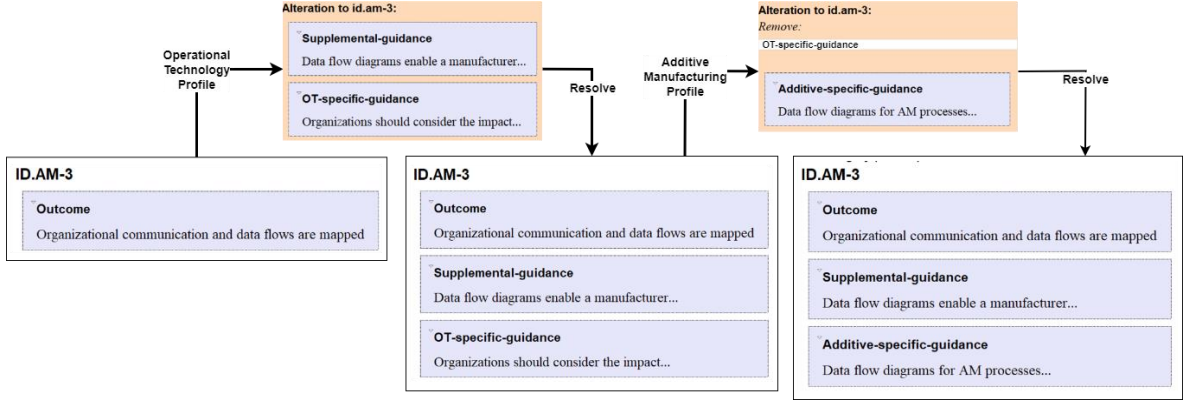

**Fig. 3 Composition of OT and additive manufacturing profiles with respect to ID.AM-3**

Discussion of such interfaces is out of scope for this paper. However, this paper asserts that such interfaces require structured digital data representation of the information content. Not to be confused with the additive manufacturing process information model discussed in Sec. 3.2, this representation must encode the structure of a CSF profile in a manner amenable to implementing authoring, publishing, and managing the content workflows for developers of guidance standards. Meeting these requirements dictate that the information content be represented in a format that is not only *machine-readable* (sufficient for rendering content for humans to navigate on output devices), but also *machine-actionable*. Machine-actionability, a step beyond machine-readability, requires representation of semantic relations between concepts [55, 56].

The Open Security Controls Assessment Language (OSCAL) provides the foundation for the desired machine-actionability. OSCAL defines a set of hierarchical structured data models for representing security information. Although OSCAL was not designed specifically to represent CSF profiles, it includes two data models that can be adapted. These are the *catalog* and *profile* models. An OSCAL catalog consists of a hierarchy of security controls. Controls can be simple statements, as is the case with CSF categories and subcategories, or they can be highly detailed, as is the case with control catalogs such as NIST Special Publication 800-53 [57]. The OSCAL profile model enables a wide variety of operations on controls such as inclusion, exclusion, merging, and modification. Thus, an OSCAL profile is capable of meeting both the composition and replacement requirements elicited from the ID.AM-3 example in Sec. 3.1.

Figure 3 graphically illustrates these requirements as they apply to the ID.AM-3 example from Sec. 3.1. The box on the lower left represents ID.AM-3 and its outcome statement as shown in Table 1. The box in the middle shows how ID.AM-3 should be presented to a reader of a CSF OT profile, with the CSF outcome plus guidance added from the Guide to OT Security all grouped together. The box on the lower right shows the desired presentation of ID.AM-3 in a CSF additive manufacturing profile, with the CSF outcome plus the Guide's supplemental guidance and the additive-specific guidance provided in Sec. 3.1. Note that it does *not* include the Guide to OT Security's OT-specific guidance.

Figure 3's upper portion shows the transformations required to produce ID.AM-3's OT profile presentation from its CSF presentation, and ID.AM-3's additive manufacturing profile presentation from its OT profile presentation. The two boxes labeled "Alteration to id.am-3:" describe how each CSF profile transforms ID.AM-3. The left hand box specifies that the OT profile adds supplemental guidance and OT-specific guidance to the CSF outcome statement. The right hand box specifies that the additive manufacturing profile removes the OT profile's OT-specific guidance and adds additive-specific guidance. The labeled arrows show the sequence of alteration and "Resolve" operations needed to produce the lower middle box from the lower left box, and the lower right box from the lower middle box. A profile resolution should happen automatically whenever a change is made to its input. Thus, modifying the guidance in the upper left box should trigger an update to the lower middle box, and modifying the guidance in the upper right box should trigger an update to the lower right box. This automatic updating ability is necessary for change management. In fact, Fig. 3's upper boxes can be compared to deltas ("diffs") in a source code management tool such as Git [58].

Let us now show how the OSCAL catalog and profile models provide the machine-actionability needed to achieve the transformations shown in Fig. 3. The source of this section's OSCAL code is [59]. The following code represents subcategory ID.AM-3 as defined in Table 1 as a control in an OSCAL catalog representation of the CSF. The code uses YAML (YAML Ain't Markup Language) [60], a human-readable syntax for representing structured data.

```yaml
- id: id.am-3
  class: subcategory
  parts:
    - name: statement
      class: outcome
      prose: Organizational communication and
        data flows are mapped
```

The preceding YAML code specifies ID.AM-3 as one item in the list of subcategories comprising category ID.AM (list items in YAML are prefaced with a dash). ID.AM-3 has three top-level key-value pairs: id, whose value is "id.am-3", class, whose value is "subcategory", and parts, whose value is a list with a single list item whose three keys represent ID.AM-3's outcome statement.

OSCAL defines a profile resolution algorithm that enables composition of profiles. This algorithm can be conceptualized as a function $r$ in Eq. (1). $r$'s input is one or more OSCAL catalogs $C_1, \cdots, C_n$ and an OSCAL profile $P$. $r$'s output is a catalog $C_R$ whose contents is the result of applying the operations specified in the profile to the input catalog(s).

$$C_R = r(C_1, \cdots, C_n, P) \qquad (1)$$

The following code represents ID.AM-3 as modified in the Guide to OT Security [7] in OSCAL as part of a CSF OT profile, which imports ID.AM-3 as-is from the CSF catalog. The code specifies two keys: control-id which identifies the control being modified, and adds whose value lists two additions to the control: general guidance and OT-specific guidance.

```yaml
- control-id: id.am-3
  adds:
    - parts:
        - name: guidance
          class: supplemental-guidance
          prose: Data flow diagrams enable a
            manufacturer...
        - name: ot-specific
          class: OT-specific-guidance
          prose: Organizations should consider the
            impact...
```

Equation (2), derived from Eq. (1), shows a resolved catalog $C_{R_{OT}}$ resulting from applying an OSCAL CSF OT profile $P_{OT}$ to an OSCAL catalog $C$ representing the CSF core. $P_{OT}$ contains the preceding code for modifying ID.AM-3. The bottom left ID.AM-3 box in Fig. 3 corresponds to $C$, the upper left "Alteration to id.am-3" box corresponds to $P_{OT}$, and the lower middle ID.AM-3 box corresponds to $C_{R_{OT}}$.

$$C_{R_{OT}} = r(C, P_{OT}) \quad (2)$$

Now let us consider the layered approach to modification of ID.AM-3 for additive manufacturing cybersecurity, as discussed in Sec. 3.1. The discussion suggested that additional general guidance from the Guide to OT Security to document data flows applies to additive manufacturing. However, the OT-specific caution regarding network-based scanning and automated mapping tools is less applicable. Therefore, a CSF additive manufacturing profile might choose to inherit the general guidance, but replace the OT-specific guidance with the stated additive manufacturing-specific guidance. The following OSCAL profile code represents this, assuming it is part of a CSF profile that imports ID.AM-3 as-is from the resolved catalog $C_{R_{OT}}$. The `removes` key specifies omission of ID.AM-3's OT-specific guidance. The `adds` key specifies that ID.AM-3 be augmented with the stated additive manufacturing-specific guidance.

```
- control-id: id.am-3
  removes:
    - by-name: ot-specific
  adds:
    - parts:
      - name: am-specific
        class: Additive-specific-guidance
        prose: Data flow diagrams for AM
          processes...
```

Equation (3) shows the result of composition, $C_{R_{Additive}}$, when applying an additive manufacturing profile $P_{Additive}$ to the resolved catalog resulting from applying $P_{OT}$. The bottom middle ID.AM-3 box in Fig. 3 corresponds to $C_{R_{OT}}$, the upper right "Alteration to id.am-3" box corresponds to $P_{Additive}$, and the lower right ID.AM-3 box corresponds to $C_{R_{Additive}}$.

$$\begin{aligned} C_{R_{Additive}} &= r(C_{R_{OT}}, P_{Additive}) \\ &= r(r(C, P_{OT}), P_{Additive}) \end{aligned} \quad (3)$$

The following code represents ID.AM-3 in $C_{R_{Additive}}$. The `parts` key contains ID.AM-3's outcome statement inherited from the CSF catalog, general guidance inherited from $C_{R_{OT}}$, and additive manufacturing-specific guidance specified in $P_{Additive}$ that replaces ID.AM-3's removed OT-specific guidance.

```
- id: id.am-3
  class: subcategory
  parts:
    - name: statement
      prose: Organizational communication and
        data flows are mapped
    - name: guidance
      class: supplemental-guidance
      prose: Data flow diagrams enable a
        manufacturer...
    - name: am-specific
      class: Additive-specific-guidance
      prose: Data flow diagrams for AM
        processes...
```

This section presented the layered approach from the viewpoint of developers of software tools that automate the production, publication, and management of guidance documents. The OSCAL catalog model was used to represent CSF outcomes, and the OSCAL profile model was used to represent the transformations required to produce CSF-based additive manufacturing guidance. A key insight is that, to achieve composability, consistency, and change management, a machine-actionable representation of a CSF profile must differ from a human reader-friendly presentation. Rather than specify a profile in a manner optimized for human readability, a profile should instead be specified as a transformation. Doing so enables profiles to be composable.

## 5 Conclusion and Future Work

Additive manufacturing technology has revolutionized manufacturing by enabling the creation of complex free-form structures that were nearly impossible to produce previously. With the increasing adoption of additive manufacturing in critical applications ranging from aerospace parts to medical devices, the need for effective security measures and risk management activities has become more pressing than ever. However, the lack of standardized security guidelines and protocols specific to additive manufacturing remains a major challenge. In effect, this creates a potential security gap that can be exploited by malicious actors seeking to gain unauthorized access to IP data or sabotage additive operations altogether. Additionally, additive manufacturing's reliance on data models and file exchanges, as well as the diversity of fabrication specifications and post-production processes, increases the potential for cyber-attacks. Such attacks can disrupt not only additive process chains, but also hardware and material supply chains.

This paper applied a layered guidance approach that provides an effective means to standardize risk-based guidance for securing additive manufacturing data assets. Leveraging guidance developed using this approach, manufacturers can create comprehensive security programs for protecting their data assets that combine existing IT and OT security guidelines with new additive-specific guidance. The result will facilitate the development of security outcomes that address the unique risks and challenges of additive manufacturing. By automating change management of security guidance documentation, standards developers can streamline their guidance publication workflows, leaving more time for risk-based analyses to better inform the additive manufacturing industry in response to new threats and technological changes.

The layered approach shown in this paper requires additional research and development to realize its full potential. Covering more of the CSF beyond ID.AM would provide additive manufacturers with a broader spectrum of security guidance. Prime areas where additive-specific guidance would be most beneficial might include supply chain risk management (ID.SC), which could draw upon research efforts such as [11] and [12], and data security (PR.DS), where the research on side-channels cited in Sec. 2 could inform the guidance. Once an optimal subset of the CSF has been covered for general additive manufacturing, this general profile could form the basis for profiles providing guidance for individual processing technologies, or even for specific materials (e.g., profiles for metal PBF and polymer PBF).

As mentioned in Sec. 3, risk assessment guidance would be of great value to the additive manufacturing industry and its customers. There is already an existing body of guidance and research on risk assessment for OT. The Guide to OT Security [7] devotes an entire section to it, with special emphasis on OT supply chain risk management and safety systems. Research works such as [17], [18], and [61] have proposed methods for measuring the risk of cyber threats to physical processes and their safety, which Lyu et al. [18] refer to as C2P (Cyber to Physical) risk. A salient characteristic of additive manufacturing with respect to C2P is that "C" can be very large. Additive manufacturing cyber-elements [11] contributing to risk might include a distributed network for collaborative CAD model design, testing, and process planning, 3D printing instructions specifying process parameters and materials, and software applications (some of which could be deployed as cloud services). Further analysis is needed to determine how much of this guidance and research is broadly applicable to additive manufacturing and, if so, what additional research is needed.

## Disclaimer

Certain commercial and third-party products are identified in this paper. Such identification does not imply recommendation or endorsement by NIST; nor does it imply that the products identified are necessarily the best available for the purpose.

## Data Availability Statement

The data and information that support the findings of this article are freely available at [59].

## Nomenclature

C2P = Cyber to Physical
CAD = Computer-aided Design
CAM = Computer-aided Manufacturing
CAPP = Computer-aided Process Planning
CAx = CAD, CAPP, and CAM
CSF = Cybersecurity Framework
DNN = Deep Neural Network
FDM = Fused Deposition Modeling
FFF = Fused Filament Fabrication
IP = Intellectual Property
IT = Information Technology
L-PBF = Laser Powder Bed Fusion
ML = Machine Learning
NDT = Non-Destructive Techniques
NIST = National Institute of Standards and Technology
OSCAL = Open Security Controls Assessment Language
OT = Operational Technology
QR = Quick Response
STEP-NC = Standard for the Exchange of Product Model Data with Numerical Control
YAML = YAML Ain't Markup Language